\documentclass[intlimits,twoside,a4paper]{article}

\usepackage{amsmath,amssymb}
\usepackage{graphicx}

\usepackage[T2A]{fontenc}
\usepackage[cp1251]{inputenc}

\usepackage[eqsecnum]{cmpj2}

\issue{2014}{17}{2}{23003}
\doinumber{10.5488/CMP.17.23003}

\title[Statistical analysis of self-similar behaviour in the shear induced melting model]
{Statistical analysis of self-similar behaviour in the shear induced melting model}
\author[I.A.~Lyashenko, V.N.~Borysiuk, N.N.~Manko]
{I.A.~Lyashenko\refaddr{label1,label2},
        V.N.~Borysiuk\refaddr{label1,label3}, N.N.~Manko\refaddr{label1}}
\addresses{
\addr{label1} Sumy State University, 2 Rimskii-Korsakov St., 40007 Sumy, Ukraine
\addr{label2} Peter Gr\"unberg Institut-1, FZ-J\"ulich, 52425 J\"ulich, Germany
\addr{label3} A.J.~Drexel Nanomaterials Institute, Drexel University, 3141 Chestnut St., Philadelphia, PA 19104, USA
}

\date{Received May 19, 2014, in final form May 24, 2014}

\begin{document}

\maketitle

\begin{abstract}
The analysis of the system behavior under the effect of the additive noises has been done using a simple model of shear melting.
The situation with low intensity of the order parameter noise has been investigated in detail, and time dependence of the order parameter
has been calculated. A distinctive feature of the obtained dependence is power-law distribution and self-similarity. The generalized
Hurst exponent of the time series has been found within multifractal detrended fluctuation analysis. It is shown that the self-similarity
of the time series increases when the noise intensity reduces.
\keywords shear melting, additive noise, self-similarity, multifractal fluctuation analysis, \\ Fokker-Planck equation
\pacs 05.70.Ln, 47.15.gm, 62.20.Qp, 68.35.Af, 68.60.--p
\end{abstract}

\section{Introduction}

It has been recently found that low-dimensional systems undergo melting during the shear deformation (i.e., ''shear melting``).
Notably, this kind
of melting is often observed in various types of colloidal crystals~\cite{first} and while
sliding along the grain boundaries under intensive plastic
deformation~\cite{SP1,metlov}, it leads to the superplasticity mode. The effects that
are outlined on bicrystals~\cite{SP3} also resemble a shear melting.
Both shear and thermodynamic melting are demonstrated by ultrathin lubricant films
clamped between atomically smooth solid
surfaces~\cite{Yosh,2lit,5lit,Aranson,Popov,2000_SolStCom_Popov}. To describe the
latter ones, synergetic~\cite{TrenieIznos2013,UFN,JtfL11}, and
thermodynamic~\cite{JtfL2,trib_system,Jtf_visc_my} models were proposed.
The paper~\cite{TrenieIznos2013} presents the situation of an ultrathin lubricant
film melting when shear stresses exceed the critical value. It has been
shown that the external load has a critical effect on the nature of the melting.

The behavior of bilayers and light-induced hydrophobic interaction between them was experimentally
investigated in paper~\cite{Izr}. The model describing the main
factors that affect the behavior of such systems was also proposed in paper~\cite{Izr}.
It quantitatively describes the experimental data received using the surface
forces apparatus (SFA). The simple model of shear melting, which can be used to describe
different types of systems, was proposed in paper~\cite{SIM}.
The dynamic phase diagram with various modes of the system behavior, depending on
the strain rate, was constructed therein. This paper received a subsequent
development in~\cite{PRE_first}, wherein the relative motion of bilayers that are
capable of forming both disordered and ordered structures, characterized by different values of
the order parameter introduced in the description, was studied using the
numerical analysis of basic equations. The additive noises were introduced in
paper~\cite{PRE_first}, but this was used rather for attracting the system
to the steady state upon any initial conditions, and it was not shown
that the noise can have a
critical effect on the character of the system behavior. The aim of the current
paper is to investigate this effect and to define the conditions of self-similar behavior
when there is no typical scale of the order parameter. The present paper also describes the
conditions at which the system demonstrates mono- or multifractal structure characterized by the
spectrum of fractal dimensions.

\section{The model and the Fokker-Planck equation}

General expression of the free energy density for the system undergoing the shear melting is as follows~\cite{PRE_first,SIM}:
\begin{equation}
F(\rho,\theta) = \frac{a_1\rho^2}{2} - \frac{b_1\rho^3}{3} + \frac{c_1\rho^4}{4} +
\frac{\alpha\rho^2}{2}\left(\frac{a_2\theta^2}{2} - \frac{b_2\theta^3}{3} + \frac{c_2\theta^4}{4}\right),
\label{free_energy}
\end{equation}
where the order parameter $\rho$ is the modulus of the density wave amplitude~\cite{SIM}.
The parameter $\theta$ is a periodic variable corresponding to the strain~\cite{SIM}.
For the solid-like state at zero deviations, $\theta=0$. For solid-like phase, $\rho > 0$
when there is an order in the system. Conversely, if the system undergoes the
shear melting, the stationary value $\rho=0$ is established. It can be noted that in
paper~\cite{PRE_first}, the energy~\eqref{free_energy} was obtained from the Brownian
dynamic simulations within the multiparticle model. Analyzing the obtained results,
the authors~\cite{PRE_first} selected constant parameters in the
potential~\eqref{free_energy}: $a_1 = 0.85$, $b_1 = 5.8$, $c_1 = 8.0$, $a_2 = 1.3644$, $b_2 = 8.7105$, $c_2 = 13.674$.
These values correspond to the first-order phase transitions
in the system.

Using the energy~\eqref{free_energy}, the system  of Landau-Khalatnikov type evolution equations can be written as~\cite{PRE_first,SIM}:
\begin{eqnarray}
\dot\rho &=& -\frac{1}{\gamma_\rho}\frac{\partial F(\rho,\theta)}{\partial \rho} + \xi_\rho(t),\label{eq1}\\
\dot\theta &=& -\frac{1}{\gamma_\theta}\frac{\partial F(\rho,\theta)}{\partial \theta} + \Omega + \xi_\theta(t),\label{eq2}
\end{eqnarray}
where white noises $\xi_q(t)$ have the following moments:
\begin{equation}
\langle\xi_q(t)\rangle = 0,\qquad \langle\xi_q(t)\xi_q(t+\tau)\rangle = 2D_q\delta(\tau),
\end{equation}
where $q=\rho,\theta$ and $D_q$ are the noise intensities.

The forcing term $\Omega$ corresponds to shearing the solid at the relative motion of
the surfaces~\cite{SIM}. The main reason of this parameter~\cite{PRE_first,SIM} is
that the equation~\eqref{eq2} is transformed into the relation~$\dot\theta=\Omega$
when there are no forces and noises. It resembles the expression $\dot\varepsilon = V/h$,
where $\varepsilon$ is a full strain in a layer,~$V$ is a shear velocity of surfaces,
$h$ is a thickness of a lubricant layer that was previously used in
describing the shear melting of the ultrathin lubricant layers. Thus,~$\Omega$ can be
presented as the motion velocity of interacting layers.

In the case $\Omega=0$, the system is described by the free energy \eqref{free_energy}.
According to the structure of the equations~\eqref{eq1}, \eqref{eq2}, the case
of~$\Omega\ne 0$ corresponds to the energy
\begin{equation}
F^\prime(\rho,\theta) = F(\rho,\theta) - \theta\Omega\gamma_\theta\,,
\label{free_energy1}
\end{equation}
which differs from the initial expression by the presence of the last component.

\begin{figure}[!t]
\centerline{
\includegraphics[width=0.5\textwidth]{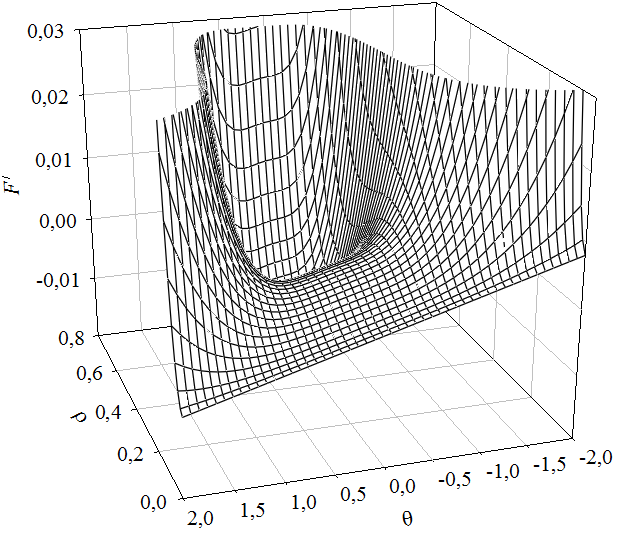}   
}
\caption{The free energy $F^\prime(\rho,\theta)$~\eqref{free_energy1} at the parameters
$\alpha = 0.17$, $\Omega = 0.08$, $\gamma_\theta = 0.05$.}
\label{fig1}
\end{figure}

The three-dimensional dependence $F^\prime(\rho,\theta)$ is shown in figure~\ref{fig1}.
The potential \eqref{free_energy1} increases in the negative region of the parameters
$\rho$ and $\theta$, and the velocity value $\Omega$ specifies the slope of the
line $F^\prime(\theta)$ at $\rho=0$. According to the type of the potential shown
in figure~\ref{fig1}, the parameter $\theta$ will monotonously increase with time in case when
there is no noise at the stationary value~$\rho=0$. However, this does not occur in
the presence of noise, and constant transitions between two attracting points are realized.
One of these points corresponds to the minimum at~$\rho\ne 0$ as shown in figure~\ref{fig1}.
In paper~\cite{PRE_first}, the effect of the noise intensity at different $\alpha$
and $\Omega$ parameters on the system behavior was studied numerically.
The analytical expressions that permit to perform such kind of analysis can be obtained.

Let us consider the Fokker-Planck equation. The system of equations \eqref{eq1} and \eqref{eq2}
can be put in correspondence with the following two-dimensional
equation~\cite{FokPl,Risken}:
\begin{eqnarray}
\frac{\partial P(\rho,\theta)}{\partial t} &=&
\frac{\partial}{\partial\rho}\left[\frac{1}{\gamma_\rho}\frac{\partial F^\prime(\rho,\theta)}{\partial\rho}P(\rho,\theta)\right]+
\frac{\partial}{\partial\theta}\left[\frac{1}{\gamma_\theta}\frac{\partial F^\prime(\rho,\theta)}{\partial\theta} P(\rho,\theta)\right]
\nonumber\\
&&+D_\rho\frac{\partial^2}{\partial\rho^2}P(\rho,\theta) + D_\theta\frac{\partial^2}{\partial\theta^2}P(\rho,\theta).
\label{FokPlank}
\end{eqnarray}
The equal relaxation times $\gamma_\rho=\gamma_\theta=\gamma$ and equal noise
intensities $D_\rho=D_\theta=D$ were considered numerically in paper~\cite{PRE_first}.
In this case, the equation~\eqref{FokPlank} can be written in a simpler form:
\begin{eqnarray}
\gamma\frac{\partial P(\rho,\theta)}{\partial t} &=&
\frac{\partial}{\partial\rho}\left[\frac{\partial F^\prime(\rho,\theta)}{\partial\rho}P(\rho,\theta)\right]+
\frac{\partial}{\partial\theta}\left[\frac{\partial F^\prime(\rho,\theta)}{\partial\theta} P(\rho,\theta)\right]
\nonumber\\
&&+\gamma D\left[\frac{\partial^2}{\partial\rho^2}P(\rho,\theta) + \frac{\partial^2}{\partial\theta^2}P(\rho,\theta)\right].
\label{FokPlank1}
\end{eqnarray}
Now, both drift coefficients exactly represent the potential derivatives~\eqref{free_energy1}.
In the stationary case $\partial P(\rho,\theta)/\partial t=0$, the solution of the
equation~\eqref{FokPlank1} provides the probability density of the type~\cite{FokPl,Risken}:
\begin{equation}
P(\rho,\theta) = C\exp\left\{-\frac{F^\prime(\rho,\theta)}{\gamma D}\right\},
\label{statFP}
\end{equation}
\begin{figure}[!b]
\centerline{
\includegraphics[width=0.43\textwidth]{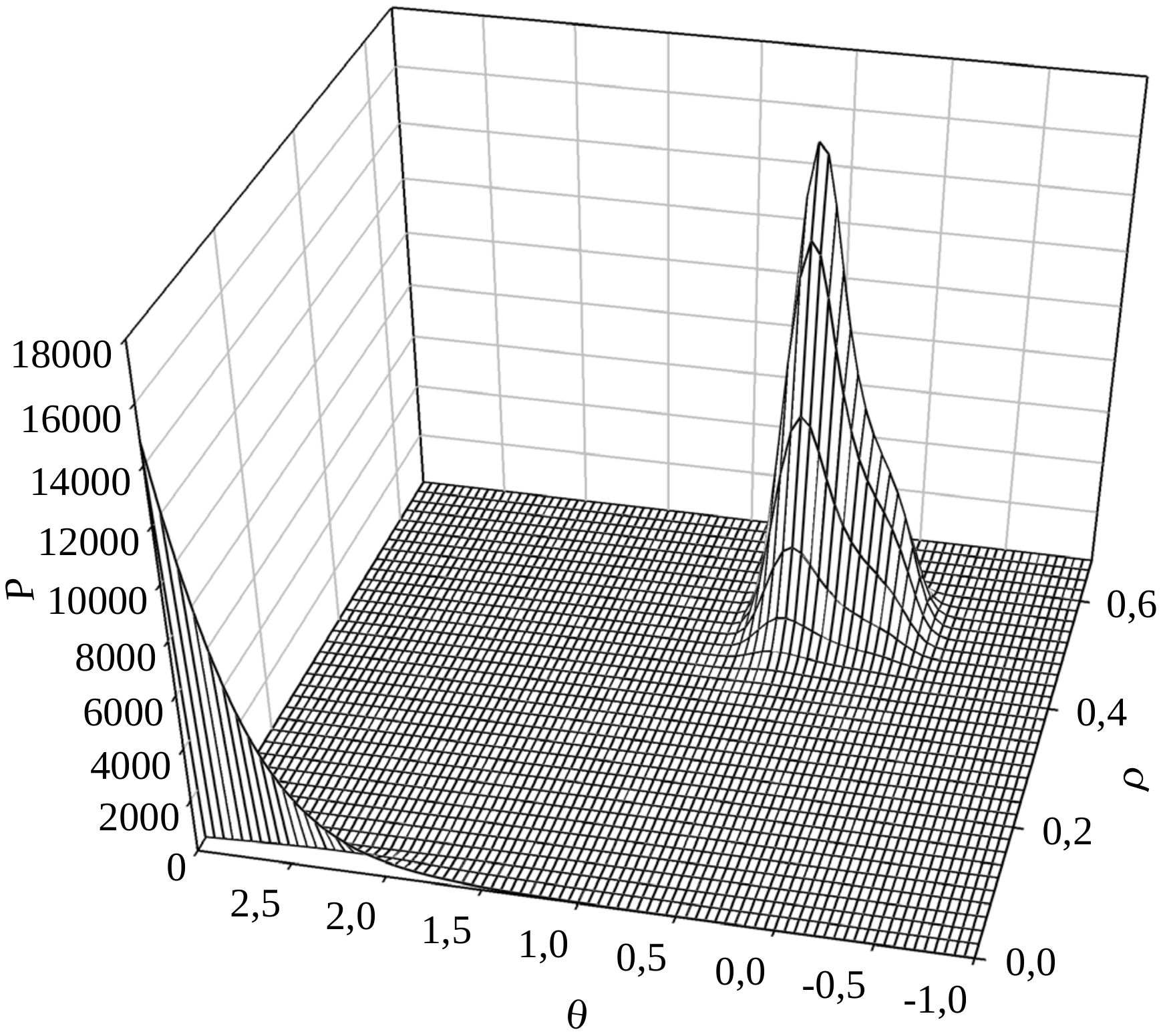}     
\hspace{3mm}
\includegraphics[width=0.43\textwidth]{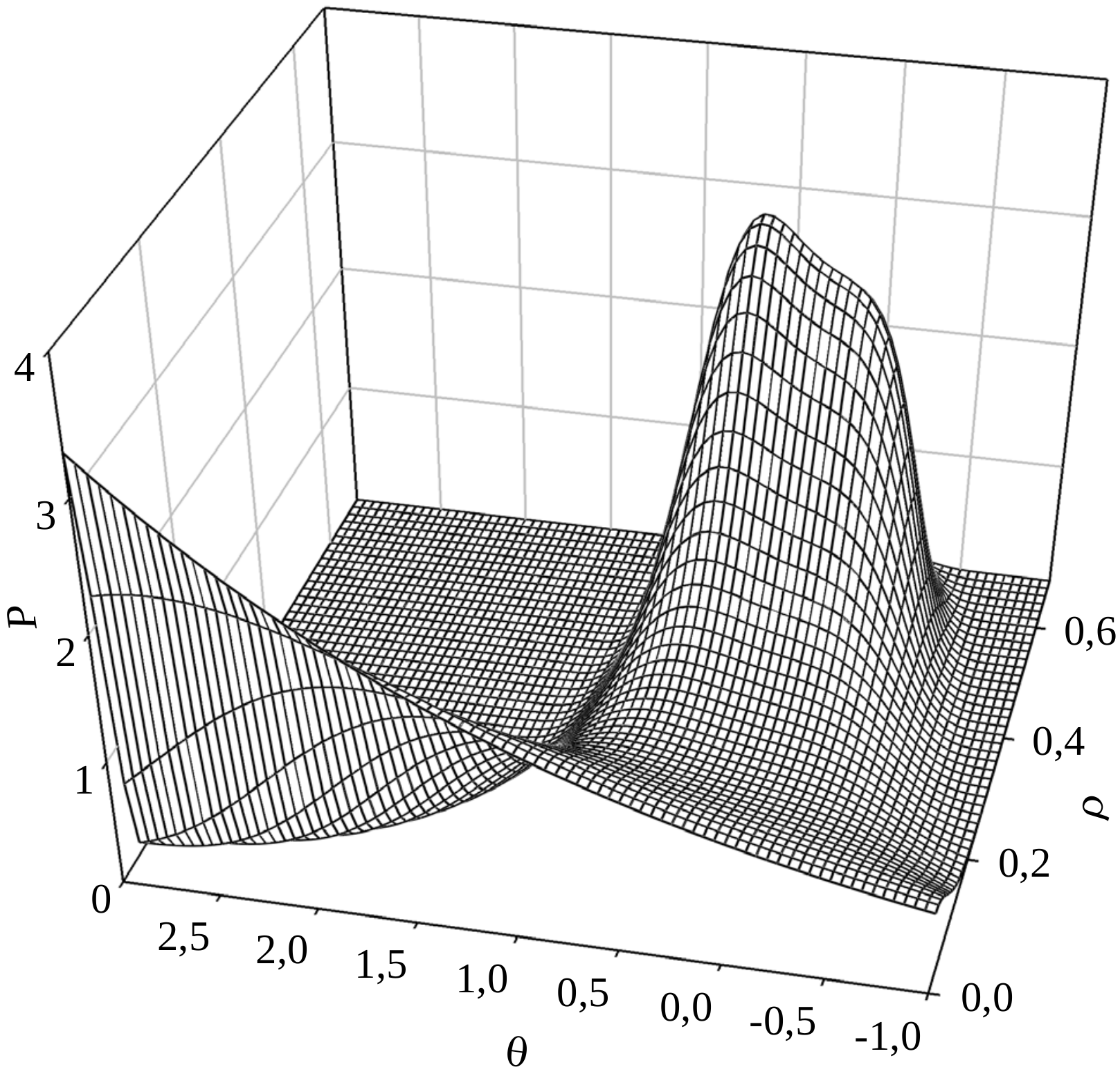}
}
\centerline{  a \hspace{0.44\textwidth} b}
\caption{The non-normalized distribution $P(\rho,\theta)$~\eqref{statFP} at the parameters of figure~\ref{fig1} and the noise intensity values:
a) $D=0.025$; b) $D=0.2$.}
\label{fig2}
\end{figure}
where $C$ is the normalization constant and $F^\prime(\rho,\theta)$ is determined by the expression \eqref{free_energy1}.
It should be noted that the coefficient $\gamma$ is also included in the energy $F^\prime(\rho,\theta)$~\eqref{free_energy1}.
The type of distribution \eqref{statFP} is shown in figure~\ref{fig2} without considering the normalization constant $C$.
The figure shows that the probability of transitions between solidlike and liquidlike
states increases with the growth of the noise intensity $D$. In the paper~\cite{PRE_first}, these
transitions have been investigated numerically in detail for different values of~$\Omega$
and~$\alpha$, and in further analysis we use fixed values of these parameters.

The numerical solution of the equations \eqref{eq1}, \eqref{eq2}, can be obtained within the Euler method~\cite{JtfL3}.
The following iterative procedure~\cite{JtfL3} corresponds to equations:
\begin{eqnarray}
\rho_{i+1} &=& \rho_i - \frac{\Delta t}{\gamma_\rho}\left[a_1\rho_i - b_1\rho_i^2 + c_1\rho_i^3 + \alpha\rho_i V(\theta_i)\right] +
\sqrt{\Delta t}W_{\rho i}\,,\label{eq1_iter}\\
\theta_{i+1} &=& \theta_i - \frac{\alpha\rho_i^2\Delta t}{2\gamma_\theta}\left(a_2\theta_i-b_2\theta_i^2+c_2\theta_i^3\right) + \Delta t\Omega +
\sqrt{\Delta t}W_{\theta i}\,,\label{eq2_iter}
\end{eqnarray}
where the potential $V(\theta_i)$ is set by the expression in brackets for the energy \eqref{free_energy}.
The random forces $W_q$ are determined according to the standard procedure~\cite{C++}
\begin{equation}
W_q = \sqrt{2D_q}\sqrt{-2\ln r_{q1}}\cos{(2\pi r_{q2})}, \qquad r_{qi} \in (0;1],
\end{equation}
and the pseudorandom numbers $r_{q1}, r_{q2}$ are characterized by a uniform distribution.

\begin{figure}[!t]
\centerline{
\includegraphics[width=0.6\textwidth]{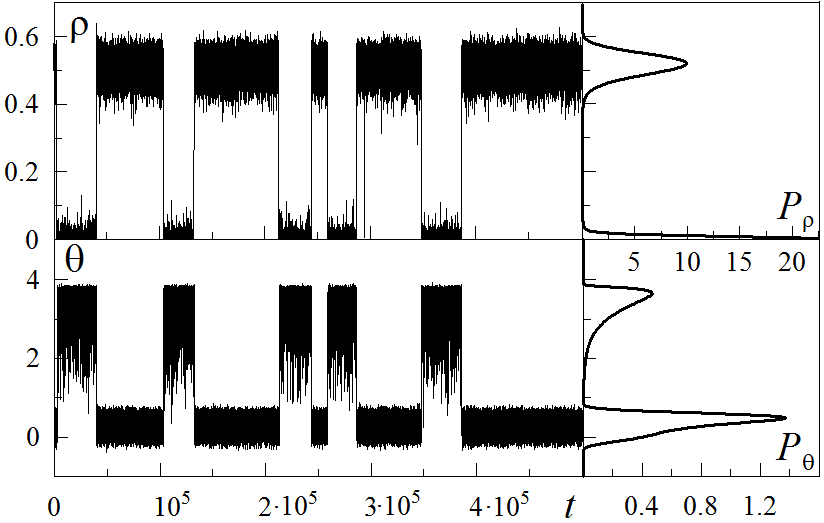}  
}
\caption{The time dependence of the absolute value of the parameter $\rho$, the value $\theta$,
and their probability densities $P_\rho(\rho)$, $P_\theta(\theta)$ at the parameters of figure~\ref{fig1}
and $\gamma_\rho = \gamma_\theta = 0.05$, $D_\rho = D_\theta = 0.025$.}
\label{fig3}
\end{figure}
\begin{figure}[!t]
\centerline{
\includegraphics[width=0.6\textwidth]{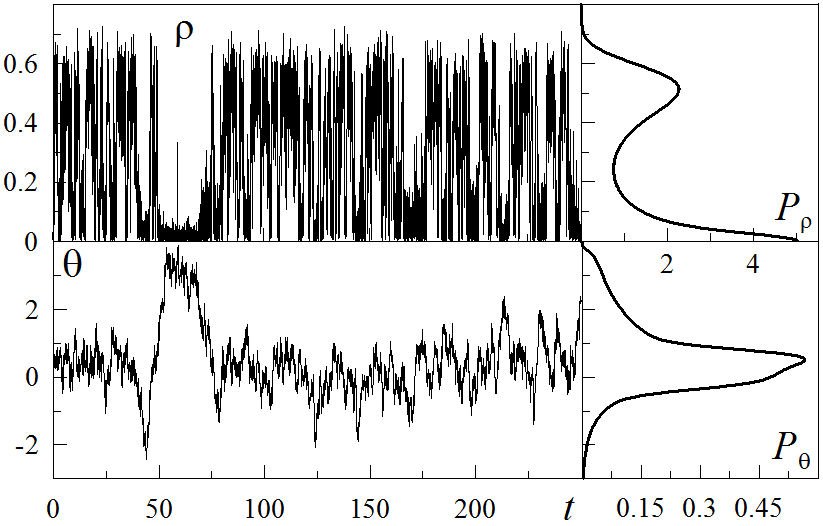}
}
\caption{The same as in figure~\ref{fig3}, but at $D_\rho=D_\theta=0.2$.}
\label{fig4}
\end{figure}

Figure~\ref{fig3} and figure~\ref{fig4} show the time dependence $\rho(t)$, $\theta(t)$,
obtained by the numerical solution of the the equations~\eqref{eq1_iter}
and \eqref{eq2_iter}\footnote{The absolute value of the parameter $\rho$ is shown
in the figures, since the region $\rho<0$ has not got a physical meaning.}.
According to the figures, regular spontaneous transitions between liquidlike~($\rho=0$)
and solidlike ($\rho\ne 0$) states occur. However, in
figure~\ref{fig3}, the probability of transitions between the ordered and disordered states is lower,
because it was built at a smaller value of the noise intensity $D$. The numerically
determined one-dimensional probability densities $P_\rho(\rho)$ and $P_\theta(\theta)$ are
shown in the right hand parts of the figures. The corresponding time series for defining the
probability densities were calculated for the time interval $t\in[0;10^6]$ with step $\Delta t = 10^{-3}$,
i.e., each series has $10^9$ points. It explains the smooth type
of the dependencies $P_\rho(\rho)$ and $P_\theta(\theta)$, which were normalized according to the conditions:
\begin{equation}
\int\limits_{0}^{+\infty}P_\rho(\rho){\rm d}\rho = 1,\qquad
\int\limits_{-\infty}^{+\infty}P_\theta(\theta){\rm d}\theta = 1,
 \end{equation}
where parameter $\rho$ was measured from zero. Thus, the areas under the probability curves in figure~\ref{fig3}
and figure~\ref{fig4} are equal to one. Note, in the figures, the curves shape for the one-dimensional
probability densities confirms the type of the two-dimensional surface shown in
figure~\ref{fig2}. The case described above was analyzed in detail in paper~\cite{PRE_first},
in which the phase diagrams with crystallization regions, liquidlike
behavior, and the region where the regular spontaneous transitions between specified states occur
(i.e., {\it stick-slip} mode), were calculated numerically based on the type
of the one-dimensional probability density $P_\rho$. Thus, the study of this issue is not
the aim of the present paper. Further, the self-similar behavior of a solidlike system will be investigated herein.

\section{The self-similar behavior}

The general Fokker-Planck equation~\eqref{FokPlank} can be solved to find out the system
behavior in general case (when relaxation times are not equal and the noise intensities are
also not equal), though this can be difficult to do since the specified equation is the second-order
equation in partial derivatives. The probability distribution at the initial stage is out of
our interest in contrast to the stationary type of distribution. It permits to replace the solution
of the Fokker-Planck equation~\eqref{FokPlank} by the numerical
analysis of the original system~\eqref{eq1_iter} and \eqref{eq2_iter}. The paper~\cite{UFN}
presents an analytical analysis of the conditions of existence of various
self-similar regimes, though equations \eqref{eq1_iter}, \eqref{eq2_iter} have a more complex structure
which complicates the analysis. Therefore, the numerical analysis will
be considered in this section.

\begin{figure}[!b]
\centerline{
\includegraphics[width=0.6\textwidth]{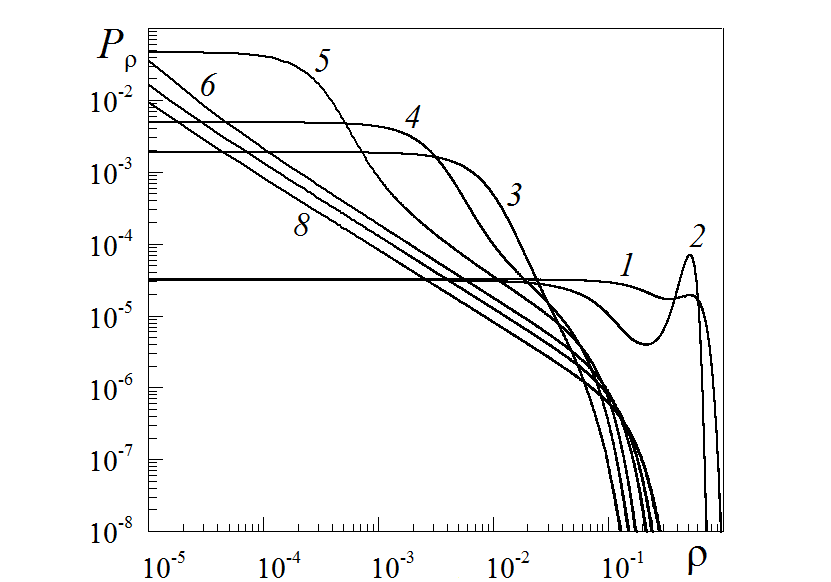}
}
\caption{The probability density $P_\rho(\rho)$, calculated at $D_\theta=10^{-2}$. The curves 1--8 correspond to the values
$D_\rho = 10^0$, $10^{-1}$, $10^{-2}$, $10^{-3}$, $10^{-5}$, $10^{-10}$, $10^{-15}$, $10^{-25}$. The curve 7
is located between the curves 6 and 8, and it is not marked in the figure.}
\label{fig5}
\end{figure}

The calculated non-normalized probability density $P_\rho(\rho)$ for different ratios between
the noise intensity values is shown in figure~\ref{fig5}, whereas the value $D_\theta$ does
not change for all curves\footnote{When obtaining the curves, the corresponding time series
were calculated for the time interval ${t\in[0;2\cdot 10^7]}$ with the
step $\Delta t=10^{-3}$. Thus, each time series has $2\cdot 10^{10}$ points. Afterwards,
the number of hits of the series values to the particular interval $\rho$ was counted.
There are $2\cdot 10^5$ points for each curve shown in the resulting figure~\ref{fig5}, i.e.,
the number of intervals on the axis $\rho$ was chosen from the value $10^{-5}$ to $3$.
The value $\rho$ did not exceed $3$ on the selected interval in the calculations.
Afterwards the number of hits into each interval was divided into the total number of points
in a series, and thus the curves fell down.}.
According to the figure we can conclude that the value of the noise intensity $D_\rho$ critically
affects the system behavior. For example, the curves 1 and 2 show the system
behavior in the two-phase region, since two maxima of probability are realized. Moreover, the
maximum at $\rho = 0$ is more expressed for the curve 1, and the curve 2 corresponds
to the case where the system is in the solidlike state ($\rho \ne 0$) for the most of time.
The two-phase region disappears when the noise intensity $D_\rho$
decreases~(curves 3--8), since only the zero maximum $P_\rho(\rho)$ is realized. The following condition is met for curves 6--8:
\begin{equation}
D_\theta \gg D_\rho\,,
\label{cond}
\end{equation}
which leads to the self-similar type of distribution density~\cite{UFN} in the region of
small order parameter values $\rho$ in this case. The fact is that the distribution
function becomes homogeneous~\cite{Ol_} in the region of small $\rho$ in the case (\ref{cond}):
\begin{equation}
P_\rho(\rho) \sim \rho^{-a},
\label{14}
\end{equation}
and corresponds to the self-similar system, for which there is no characteristic parameter scale
$\rho$  (${0<a<1}$ is an index of distribution specifying the slope of
distribution on a linear region)~\cite{Amit}. Note that the value $a=1.5$ corresponds to
a self-organized criticality mode. The case presented in figure~\ref{fig5}
shows the value $a\approx 1$.

\begin{figure}[!t]
\centerline{
\includegraphics[width=0.9\textwidth]{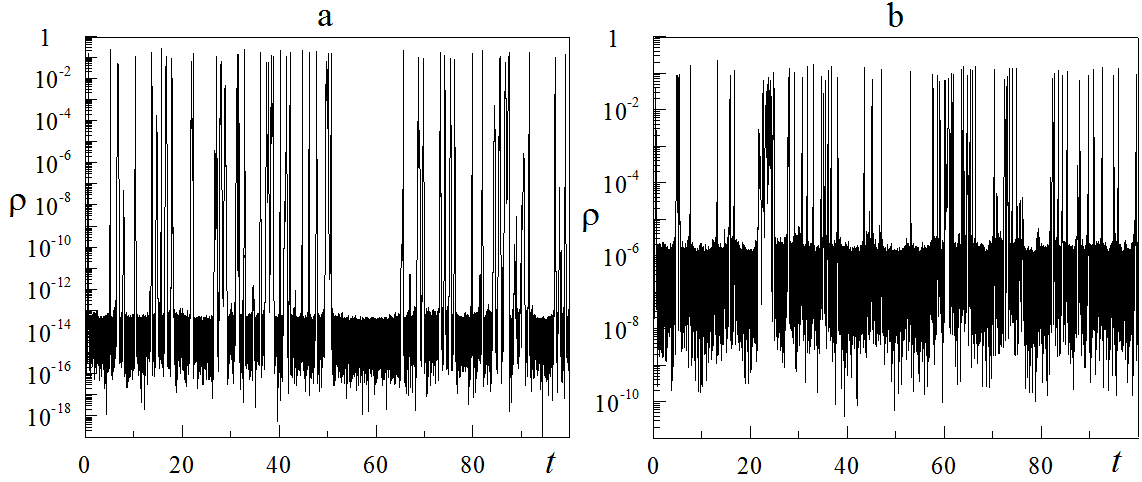}
}
\caption{The time dependence $\rho(t)$ corresponding to the parameters of figure~\ref{fig5}: a)~${D_\rho = 10^{-25}}$; b)~${D_\rho = 10^{-10}}$.}
\label{fig6}
\end{figure}

The figure~\ref{fig6} shows the time series $\rho(t)$ that corresponds to parameters of curves 8 and 6 in figure~\ref{fig5}.
The dependencies are presented in the logarithmic scale in order to show that the
self-similar behavior (the sharp increase of the order parameter values is observed
on both small
and large scale) is established in some range of the order parameter values. A smaller noise
intensity $D_\rho$ corresponds to the dependency in figure~\ref{fig6}~(a) leading to the
self-similar behavior on a larger scale range which is clearly seen from the dependency.
Figure~\ref{fig6}~(b), for which the distribution function in some interval $\rho$ is also
self-similar according to figure~\ref{fig5}, demonstrates the changes of the parameter~$\rho$
on the smaller scale range. The time series at the same noise intensities
$D_\rho=D_\theta$ are shown in figure~\ref{fig3} and in figure~\ref{fig4}, for which the self-similar
behavior is not observed. Thus, the exponential type of a distribution
function $P_\rho(\rho)$ is limited by a minimal value of the parameter~$\rho$,
which reduces with a decrease of the value~$D_\rho$.

\section{Statistical analysis of time series}

As  it was mentioned in the previous section, the characteristic feature of the
time dependence of the order parameter obtained for the noise intensity
values $D_\rho = 10^{-2}$, $10^{-3}$, $10^{-5}$, $10^{-10}$, $10^{-15}$, $10^{-25}$
(the curves 3--8 in figure~\ref{fig5}) is the presence of the power law of the distribution
in a limited range, and, as a consequence, the self-similar structure of the time series.
It should be noted that statistical parameters calculated in the previous section do not
provide a full information about the time series behavior. Thus, the absence of the
characteristic scale at different time periods is not considered
while calculating the standard statistical parameters for the self-similar time series.
Such a feature can be considered within the framework of the \emph{scaling} analysis.
One of the possible techniques of detecting the local properties of the time dependency is
the method of multifractal detrended fluctuation analysis~\cite{kantel} that permits to
examine the time series of various nature~\cite{v1,v2}.

The algorithm of the mentioned method has the following steps (see an original description
in the paper~\cite{kantel}). First, the fluctuation profile
\begin{equation}
\label{met1} y(i) = \sum\limits_{k = 1}^i {\left[ {x(k) - \bar {x}}
\right]} \,,
\end{equation}
measured from the average value $\bar {x}$, is calculated  from the considered series
$x(k)$, $k = 0,1,2,\ldots,N$. The obtained values $y(i)$ are separated by disjoint segments of
length $s$, the number of which is equal to the integer value $N_s = [N / s]$.
As a series length $N$ is not always a multiple of the selected scale $s$, the last section has less
points than $s$ in general case. To consider this residue, it is necessary to repeat
the separation into segments starting from the opposite end of the series. As a result, the total
number of segments having length $s$ is $2N_s $.

Since the changes of the random value $y(i)$ occur close to the value $y_\nu (i) \ne 0$ due
to the definite trend of the series evolution, the local trend $y_\nu (i)$ should be
found for each $2N_s$ segments. It is convenient to use the least squares method presenting
the trend $y_\nu (i)$ as a polynomial of a certain degree to ensure that interpolation
error should not exceed the specified limit. The next step is to calculate of the fluctuation function
\begin{equation}
\label{met2} F^2(\nu ,s) = \frac{1}{s}\sum\limits_{i = 1}^s {\left\{
{y\left[ {\left( {\nu - 1} \right)s + i} \right] - y_\nu (i)}
\right\}} ^2,
\end{equation}
for the segments $\nu = 1,\ldots,N_s $, going in the forward direction, and the corresponding value
\begin{equation}
\label{met3} F^2(\nu ,s) = \frac{1}{s}\sum\limits_{i = 1}^s {\left\{
{y\left[ {N - \left( {\nu - N_s } \right)s + i} \right] - y_\nu (i)}
\right\}} ^2,
\end{equation}
for the reverse sequence $\nu = N_s + 1,\ldots,2N_s $.

The next step presents a generalization of the fluctuation function
\begin{equation}
\label{met4} F_q (s) = \left\{ {\frac{1}{2N_s }\sum\limits_{\nu =
1}^{2N_s } {\left[ {F^2(\nu ,s)} \right]^{q / 2}} } \right\}^{1 / q},
\end{equation}
via raising the expressions (\ref{met2}), (\ref{met3}) to the power $q$ and the subsequent
averaging over all segments. Since the equation (\ref{met4}) has an uncertainty at $q = 0$,
the limit expression should be used instead of it
\begin{equation}
\label{met5} F_{0}(s) = \exp{\frac{1}{4N_s}\sum\limits_{\nu = 1}^{2N_s }\ln [ F^{2}(\nu ,s)]}.
\end{equation}
By changing the time scale $s$ at the fixed parameter $q$, the $F_q (s)$ dependence should
be presented in the double logarithmic coordinates. Reducing the analyzed series  to
a self-similar set showing long-range correlations, the fluctuation function $F_q (s)$ can be
presented as an exponential dependency
\begin{equation}
F_q (s) \propto s^{h(q)},
\label{met6}
\end{equation}
with the generalized Hurst exponent $h(q)$, the value of which is determined by the parameter $q$.
The definitions (\ref{met4}), (\ref{met6}) show that this parameter is reduced
to the classical Hurst exponent~$H$ at $q = 2$. If the fluctuation function $F^2(\nu ,s)$ is the
same for all segments $\nu $ and the generalized Hurst exponent $h(q) = H$ does
not depend on the parameter $q$, the time series corresponds to a monofractal set. For multifractal
series at positive $q$, the main contribution to the function $F_q (s)$ is
provided by the segments $\nu$ showing large deviations $F^2(\nu ,s)$, and the segments with a small
fluctuation values  $F^2(\nu,s)$ dominate at negative $q$. As a result, it
can be concluded that the generalized Hurst exponent $h(q)$ describes the segments displaying
small fluctuations at negative values $q$, and large fluctuations at positive
values~\cite{kantel,feder}.

It should be noted that if the size of the segments increases to $s > N / 4$ during the
implementation of the above-mentioned algorithm, the function $F_q (s)$ loses the
statistical informative value due to the smallness of the number of segments $N_s < 4$ used
in the procedure of averaging. Thus, the realization of the specified procedure
presupposes an exception of large segments $(s > N / 4)$ on the one hand, and the small ones $(s < 10)$ on the other hand.

The standard representation of the time series scaling properties presupposes the transition
from the Hurst exponent $h(q)$ to the mass index $\tau (q)$ and the spectral
function $f(\alpha )$, which are both the main characteristics of multifractals~\cite{kantel,feder}:
\begin{equation}
\label{met7} \tau (q) = qh(q) - 1,
\end{equation}
\begin{equation}
\label{met8} f(\alpha ) = \alpha q(\alpha ) - \tau (q(\alpha )).
\end{equation}
Herein the value $q(\alpha )$ is determined by the condition ${\tau }'(q) = \alpha $, where
the prime symbol means a differentiation by an argument. The dependency $\tau(q)$
has a linearly increasing form at $|q|\gg 1$ with the curved section near $q = 0$, which
provides the deceleration of the mass index $\tau $ growth with the parameter $q$ increasing.
The spectral function $f(\alpha )$ determines a monofractal set having the dimensions $\alpha$,
which forms the structure under investigation, wherein the relative number of
monofractals with dimension $\alpha $, within the segments with the size $l$, covering this set,
is defined by the relation $N(\alpha ) \sim l^{ - f(\alpha )}$. According to this
definition, $f(\alpha )$ represents the number of different monofractals in the set.
Thus, the spectral function $f(\alpha)$ for the monofractal set is $\delta$-shaped with a
single value of the fractal dimension $\alpha$~\cite{feder}.

\begin{figure}[!t]
\centerline{
\includegraphics[width=0.47\textwidth]{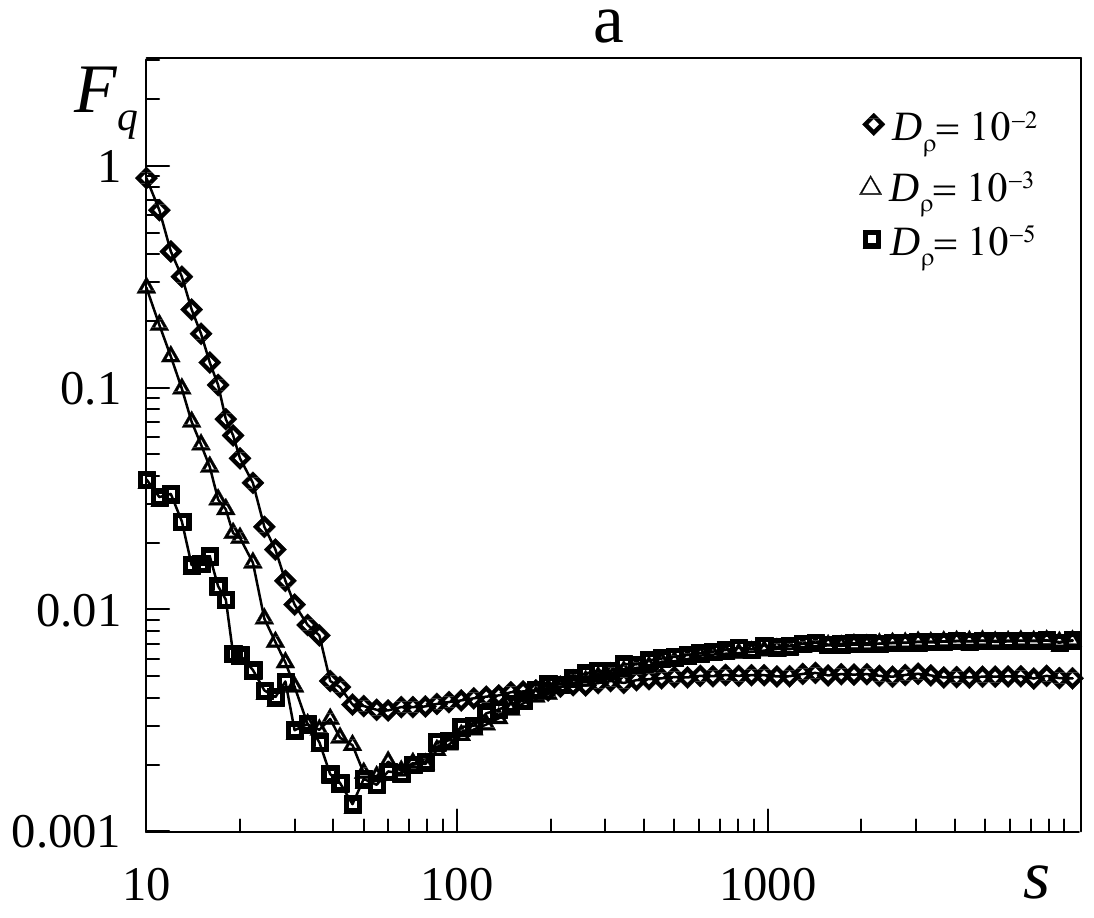}
\
\includegraphics[width=0.47\textwidth]{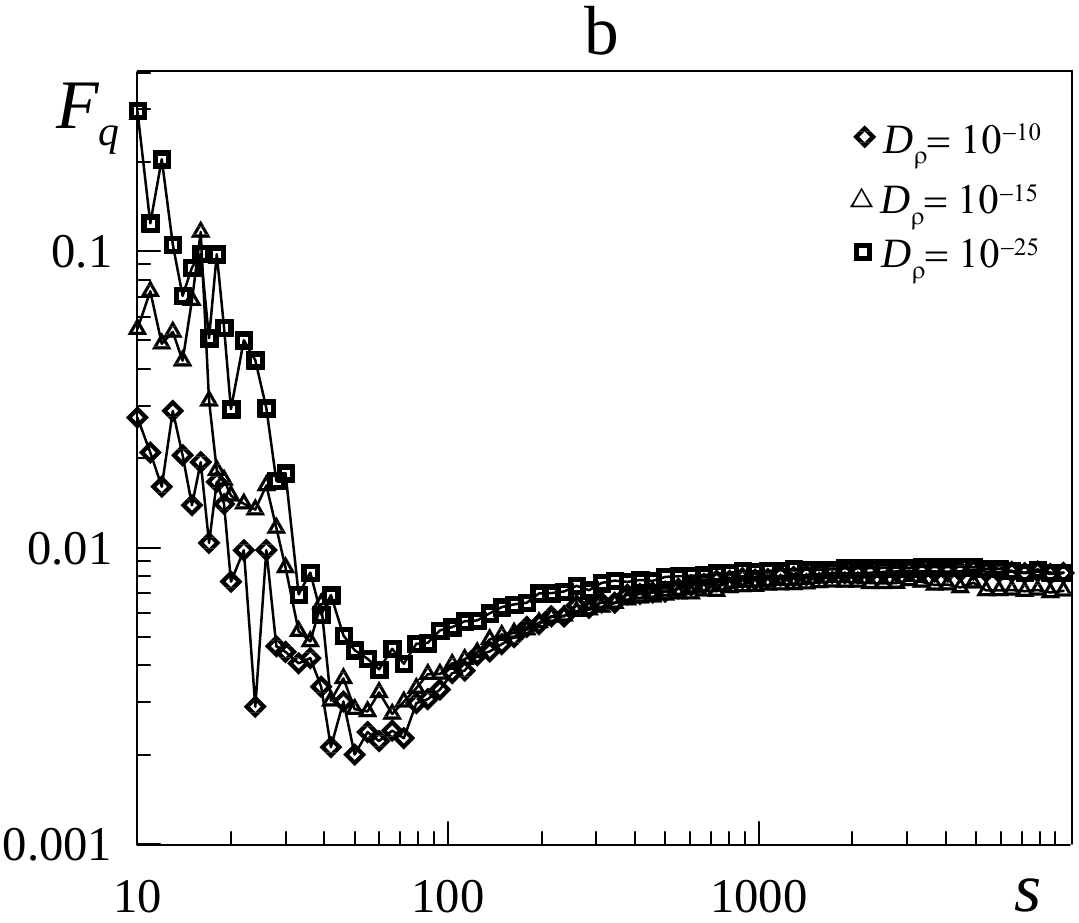}
}
\caption{The dependence (\ref{met6}), built in double logarithmic axes at the deformation
value $q=2$ for the series with the values a) $D_\rho = 10^{-2}$, $10^{-3}$, $10^{-5}$
 and b) $D_\rho = 10^{-10}$, $10^{-15}$, $10^{-25}$ .}
\label{fig7}
\end{figure}

\begin{figure}[!t]
\centerline{
\includegraphics[width=0.5\textwidth]{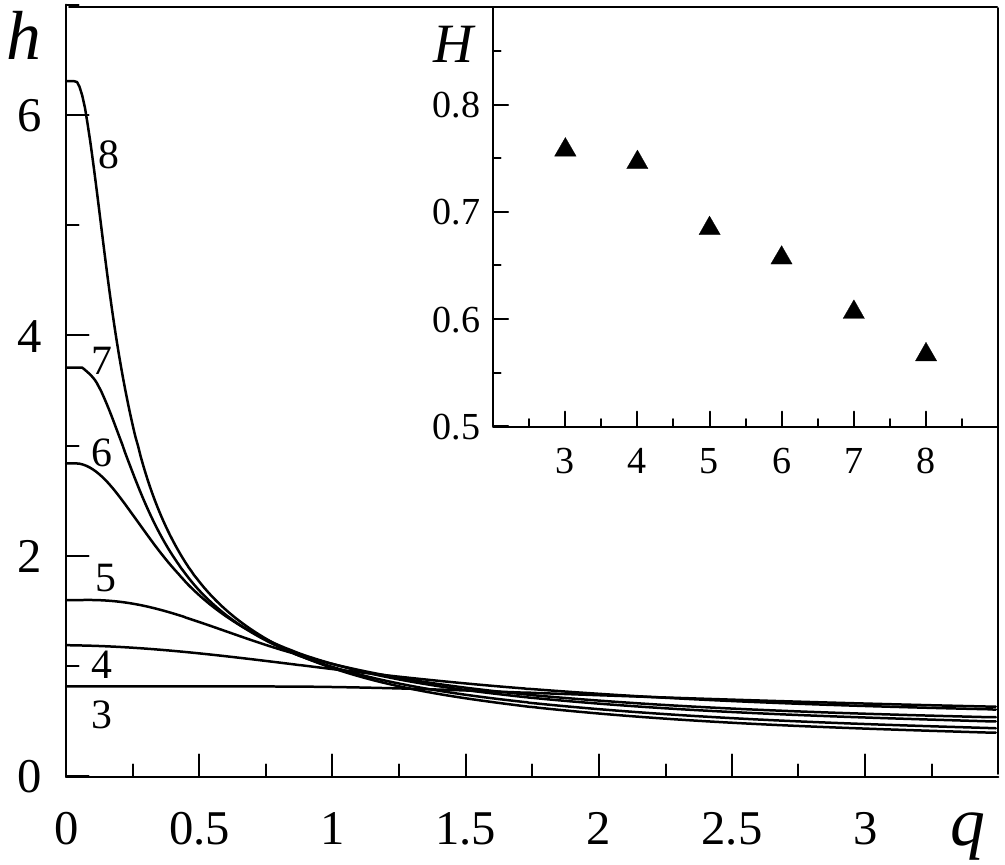}
}
\caption{The generalized Hurst exponent $h(q)$ for time series having noise intensity
$D_\rho = 10^{-2}$, $10^{-3}$, $10^{-5}$, $10^{-10}$, $10^{-15}$, $10^{-25}$ (the curves
3--8, respectively), and changes of the classical Hurst exponent $H$ for specified curves.}
\label{fig8}
\end{figure}

\begin{figure}[!b]
\centerline{
\includegraphics[width=0.5\textwidth]{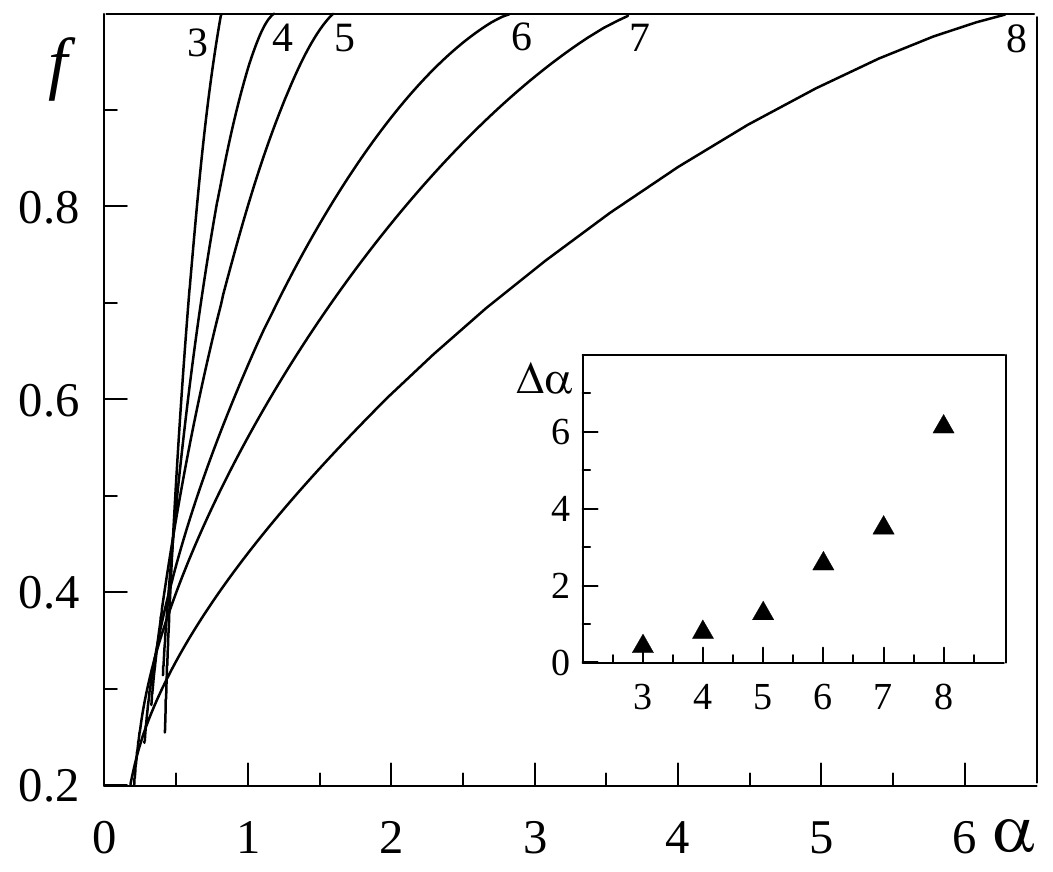}
}
\caption{The spectral function $f(\alpha)$ for time series having the noise intensity $D_\rho = 10^{-2}$, $10^{-3}$, $10^{-5}$, $10^{-10}$, $10^{-15}$, $10^{-25}$~(the curves
3--8 respectively), and the spread width of the multifractal spectrum $\Delta\alpha$.}
\label{fig9}
\end{figure}

The typical dependence (\ref{met6}) for the series having the noise intensities $D_\rho = 10^{-2}$, $10^{-3}$, $10^{-5}$, $10^{-10}$, $10^{-15}$,
$10^{-25}$ at the deformation value $q=2$ is
shown in figure~\ref{fig7}. The cubic polynomial was used in the detrending procedure (\ref{met2})--(\ref{met3})
as in the original paper~\cite{kantel} for the  multifractal series having a power-law distribution function.
The use of the fitting polynomial of a higher degree does not lead to
any change in the final results since the trend of high ($>2$) order does not present the original series.
The dependence (\ref{met6}) built in double logarithmic axes, has a strongly expressed linear section with scale values $50<s<500$, and
therefore, can be used to calculate the parameter $h(q)$.
The linear interpolation of the equation (\ref{met6}) calculated for the examined series,
within the specified interval of the changes of the scale $s$ at values of the deformation
parameter $0 \leqslant q \leqslant 3.5$, leads to the dependence $h(q)$ shown in figure~\ref{fig8}, where
the dependence of the classical Hurst exponent $H$ on the number~(noise intensity)
of the corresponding series is also shown in the additional panel.
The spectral function $f(\alpha)$ was also calculated using the equations (\ref{met7}) and (\ref{met8}) for the examined series.
The result of the performed calculations is shown in figure~\ref{fig9}.
The dependencies, presented in figure~\ref{fig8} and~\ref{fig9} show that the reduction of the noise
intensity $D_\rho$ leads to a significant complication of the order parameter
time series dynamics, expressed in the increase of the range of the values of the generalized
Hurst exponent $h(q)$ and the multifractal spectrum function~(\ref{met8}).
The increase of the spread $\Delta\alpha$ is caused by the growth of the subsets
(the so-called monofractals) number $N(\alpha)$ with the Holder parameter $\alpha$ in the time
series segments of the length $s$. This situation means that the number of the statistically
different scenarios of the system evolution increases. Furthermore, with the reduction
of $D_\rho$, the Hurst exponent $H$ approaches the value $H\approx 0.5$ which, as it is known,
corresponds to the absolutely random sequence~\cite{feder}. Thus, the complication
of the time series structure makes their further behavior unpredictable. The presence of
extremely large ejections of the order parameter values corresponds to this situation,
as shown in figure~\ref{fig6}.

\section{Conclusions}

The paper describes the model of shear melting observed in colloidal crystals of various types.
This model describes the relative motion of the pair of
interacting layers characterized by different values of the order parameter. It has been
found that the external additive noise, taken into consideration, has a critical
effect on the character of the system behavior, and the probability of transition between solidlike
and liquidlike states increases with an increase of the noise intensity.
Likewise, the case is considered when the intensity of one of the noises is much higher than the intensity
of the other.  It has been shown that the self-similar
behavior of a solidlike system is established in this case, i.e., the distribution density function for
the order parameter time series becomes of the power law form in the
limited range. This characteristic feature of the time series was detected while calculating
 standard statistical parameters of the series. A more detailed
information on the local properties of the time dependence was obtained using the method
of multifractal detrended fluctuation analysis which permits to examine
the time series of various nature. Thereafter, the conditions under which the system demonstrates
the monofractal or multifractal behavior characterized by a spectrum of
fractal dimensions were investigated.

\section*{Acknowledgements}

The paper was carried out under financial support of the Fundamental Researches State Fund of Ukraine
in the framework of Grant of President of
Ukraine GP/F49/044 ``Thermodynamic theory of slippage on grain boundaries in problem of nanostructured
metals superplasticity'' (No.~0113U007248).
Some results were partially obtained at support of the Ministry of Education and Science of Ukraine
within the framework of the project ``Modelling of friction of
metal nanoparticles and boundary liquid films which interact with atomically flat surfaces''~(No.~0112U001380).
The work was partially carried out during the stay of I.A.L. in the Forschungszentrum J\"ulich (Germany)
with a research visit due to the invitation by N.J. Persson.

\ukrainianpart

\title{Статистичний аналіз самоподібної поведінки в моделі зсувного плавлення }
\author{Я.О. Ляшенко\refaddr{label1,label2}, В.М. Борисюк\refaddr{label1,label3}, Н.М. Манько\refaddr{label1}}
\addresses{
\addr{label1} Сумський державний університет, вул. Римського-Корсакова, 2, 40007 Суми, Україна
\addr{label2} Інститут Петера Грюнберга, Дослідницький центр Юліху, D-52425 Юліх, Німеччина
\addr{label3} Інститут наноматеріалів А. Дж. Дрекселя, Університет Дрекселя, вул. Честнат, 3141, Філадальфія, Пенсильванія, США
}

\makeukrtitle

\begin{abstract}
\tolerance=3000%
Використовуючи модель зсувного плавлення, проведено аналіз поведінки системи під впливом адитивних шумів.
Детально розглянуто ситуацію, коли інтенсивність шуму параметра порядку приймає мале значення. У цьому випадку знайдено часову залежність
параметра порядку, характерною особливістю якої є степенева функція густини розподілу і самоподібність. За допомогою методу мультифрактального
флуктуаційного аналізу розраховано модифікований показник Херста для часових рядів. Показано, що самоподібні властивості рядів стають більш
вираженими зі зменшенням інтенсивності шуму.

\keywords зсувне плавлення, адитивний шум, самоподібність, мультифрактальний флуктуаційний аналіз, рівняння Фоккера-Планка
\end{abstract}

\begin{thebibliography}{00}
\bibitem{first} Ackerson~B.J., Clark~N.A., Phys. Rev. Lett., 1981, \textbf{46}, 123; \bibdoi{10.1103/PhysRevLett.46.123}.
\bibitem{SP1} Kaibyshev~O.A., Superplasticity of Alloys, Intermetallides and Ceramics, Springer-Verlag, Berlin, 1992.
\bibitem{metlov} Metlov~L.S., Myshlyaev~M.M., Khomenko~A.V., Lyashenko~I.A., Tech. Phys. Lett., 2012, \textbf{38}, 972; \bibdoi{10.1134/S1063785012110107} [Pis’ma Zh. Tekh. Fiz., 2012, \textbf{38}, 28 (in Russian)].
\bibitem{SP3} Gleiter~H., Phys. Status Solidi (b), 1971, \textbf{45}, 9; \bibdoi{10.1002/pssb.2220450102}.
\bibitem{Yosh} Yoshizawa~H., Israelachvili~J., J. Phys. Chem., 1993, \textbf{97}, 11300; \bibdoi{10.1021/j100145a031}.
\bibitem{2lit} Smith~E.D., Robbins~M.O., Cieplak~M., Phys. Rev. B, 1996, \textbf{54}, 8252; \bibdoi{10.1103/PhysRevB.54.8252}.
\bibitem{5lit} Carlson~J.M., Batista~A.A., Phys. Rev. E, 1996, \textbf{53}, 4153; \bibdoi{10.1103/PhysRevE.53.4153}.
\bibitem{Aranson} Aranson~I.S., Tsimring~L.S., Vinokur~V.M., Phys. Rev. B, 2002, \textbf{65}, 125402; \bibdoi{10.1103/PhysRevB.65.125402}.
\bibitem{Popov} Popov~V.L., Tech. Phys., 2001, \textbf{46}, 605; \bibdoi{10.1134/1.1372955} [Zh. Tekh. Fiz., 2001, \textbf{71}, 100 (in Russian)].
\bibitem{2000_SolStCom_Popov} {Popov~V.L., Solid State Commun., 2000, \textbf{115}, 369; \bibdoi{10.1016/S0038-1098(00)00179-4}.}
\bibitem{JtfL11} Khomenko~A.V., Lyashenko~I.A., Tech. Phys., 2005, \textbf{50}, 1408; \bibdoi{10.1134/1.2131946} [Zh. Tekh. Fiz., 2005, \textbf{75}, 17 (in Russian)].
\bibitem{TrenieIznos2013} Lyashenko~I.A., Manko~N.N., J. Frict. Wear, 2013, \textbf{34}, 38;
\bibdoi{10.3103/S1068366613010091} [Trenie i Iznos, 2013, \textbf{34}, 50 (in Russian)].
\bibitem{UFN} Khomenko~A.V., Lyashenko~I.A., Physics-Uspekhi, 2012, \textbf{55}, 1008; \bibdoi{10.3367/UFNe.0182.201210f.1081} [Usp. Fiz. Nauk, 2012, \textbf{182}, 1081 (in Russian); \bibdoi{10.3367/UFNr.0182.201210f.1081}].
\bibitem{JtfL2} Lyashenko~I.A., Tech. Phys., 2011, \textbf{56}, 869; \bibdoi{10.1134/S1063784211060168} [Zh. Tekh. Fiz., 2011, \textbf{81}, 125 (in Russian)].
\bibitem{trib_system} Lyashenko~I.A., Tech. Phys., 2011, \textbf{56}, 701; \bibdoi{10.1134/S1063784211050227} [Zh. Tekh. Fiz., 2011, \textbf{81}, 115 (in Russian)].
\bibitem{Jtf_visc_my} Lyashenko~I.A., Tech. Phys., 2013, \textbf{58}, 1016; \bibdoi{10.1134/S106378421307013X} [Zh. Tekh. Fiz., 2013, \textbf{83}, 87 (in Russian)].
\bibitem{Izr} Donaldson~S., Lee~T., Chmelka~B., Israelachvili~J., PNAS, 2011, \textbf{108}, 15699; \bibdoi{10.1073/pnas.1112411108}.
\bibitem{SIM} Lahiri~R., Ramaswamy~S., Phys. Rev. Lett., 1994, \textbf{73}, 1043; \bibdoi{10.1103/PhysRevLett.73.1043}.
\bibitem{PRE_first} Das~M., Ananthakrishna~G., Ramaswamy~S., Phys. Rev. E, 2003, \textbf{68}, 061402; \bibdoi{10.1103/PhysRevE.68.061402}.
\bibitem{FokPl} Horsthemke~W., Lefever~R., Noise-Induced Transitions: Theory and Applications in Physics, Chemistry, and Biology,
Springer-Verlag, New York, 1984.
\bibitem{Risken} Risken~H., The Fokker-Planck Equation, Springer, Berlin, 1989.
\bibitem{JtfL3} Lyashenko~I.A., Tech. Phys., 2012, \textbf{57}, 17; \bibdoi{10.1134/S1063784212010173} [Zh. Tekh. Fiz., 2012, \textbf{82}, 19 (in Russian)].
\bibitem{C++} Press~W.H., Teukolsky~S.A., Vetterling~W.T., Flannery~B.P.,
Numerical Recipes in C: the Art of Scientific Computing, 2nd ed., Cambridge University Press, New York, 1992.
\bibitem{Ol_} Olemskoi~A.I., Physics-Uspekhi, 1998, \textbf{41}, 269; \bibdoi{10.1070/PU1998v041n03ABEH000377} [Usp. Fiz. Nauk, 1998, \textbf{168}, 287 (in Russian);
    \bibdoi{10.3367/UFNr.0168.199803c.0287}].
\bibitem{Amit} Amit~D.J., Field Theory, the Renormalization Group, and Critical Phenomena, McGraw-Hill, Inc., New York, 1978.
\bibitem{kantel} Kantelhardt~J.W., Zschiegner~S.A., Koscielny-Bunde~E., Havlin~S., Bunde~A., Stanley~H.E., Phys. A, 2002, \textbf{316}, 87; \bibdoi{10.1016/S0378-4371(02)01383-3}.
\bibitem{v1} Olemskoi~A., Shuda~I., Borisyuk~V., Europhys. Lett., 2010, \textbf{89}, 50007; \bibdoi{10.1209/0295-5075/89/50007}.
\bibitem{v2} Pogrebnjak~A.D., Borisyuk~V.N., Bagdasaryan~A.A., Condens. Matter Phys., 2013, \textbf{16}, 33803; \bibdoi{10.5488/CMP.16.33803}.
\bibitem{feder} Feder~J., Fractals, Plenum Press, New York, London, 1988.

\end{thebibliography}
\end{document}